\documentstyle[prl,aps,epsf,floats]{revtex}
\begin{document}
\draft

\twocolumn[\hsize\textwidth\columnwidth\hsize\csname @twocolumnfalse\endcsname

\title{
Current hysteresis and memory effect in a molecular quantum dot with 
strong electron-vibron interaction
}
\author{
A.S. Alexandrov$^{1,2}$  and  A.M.  Bratkovsky$^1$
}
\address{$^1$Hewlett-Packard Laboratories, 1501 Page Mill 
Road, 1L, Palo Alto, California 94304\\
$^2$Department of Physics, Loughborough University,
Loughborough LE11 3TU, United Kingdom}
\date{December 16, 2002}
\maketitle

\begin{abstract}
Theory of current hysteresis for tunneling through a molecular quantum dot (MQD) with 
strong electron-vibron interactions and ${\it attractive}$ electron-electron
correlations is developed. The dot is modeled as 
a $d$-fold degenerate energy level weakly coupled to the leads. The effective
attractive interaction between polarons in the dot results in  a "switching" phenomenon
in the current-voltage characteristics when $d>2$, in agreement with the results for 
the phenomenological negative-$U$ model. The degenerate MQD with strong electron-vibron 
coupling has two stable current states in certain interval of the bias voltage below some 
critical temperature.

\end{abstract}

\vskip2pc]

\narrowtext

\section{Introduction}

Strongly correlated transport through mesoscopic systems with repulsive
electron-electron interactions has received considerable interest in the
past (see, for example, \cite{kas,but,and,lee,sha,lik,meir,her}), and
continues to be the focus of intense experimental and theoretical
investigation \cite{par,het}. The Coulomb interaction suppresses tunneling
for certain range of applied voltages, leading to what is commonly called
the Coulomb blockade. There is now a growing interest in molecular nanowires
and quantum dots used as the ``transmission lines'' \cite{lehn90,tour00} and
active molecular elements \cite{mark98,pat99} in molecular-scale electronics 
\cite{mark98}. 

A few experimental studies \cite{pat99} provide evidence for
a molecular switching effect, when the current-voltage (I-V) characteristics
show two branches with high and low current for the same voltage. The effect
exists in simple molecules too\cite{exp}. It is important to identify the
actual mechanism of switching. 

Recently we have proposed a negative$-U$
Hubbard model of a $d$-fold degenerate quantum dot, with an intrinsic
non-retarded switching mechanism when $d>2$ \cite{alebrawil}. We argued that
the {\em attractive} electron correlations could be caused by a strong
electron-phonon (vibron) interaction in the molecule, and/or by the valence
fluctuations.

It has been recently demonstrated that the low-bias conductance of
molecules is dominated by resonant tunneling through coupled
electronic and vibration levels \cite{zhit02}. 
Different aspects of the electron-phonon/vibron (e-ph) interaction effect on the
tunneling through molecules and quantum dots (QD) have been studied by
several authors \cite{win2,li,kan,erm,ven,fis,lun,gog}. In particular,
Glazman and Shekhter, and later Wingreen {\it et al.}\cite{win2} presented
the exact resonant-tunneling transmission probability fully taking into
account the e-ph interaction on a nondegenerate resonant site. Phonons
produced transmission sidebands but did not affect the integral
transmission probability. Li, Chen and Zhou \cite{li} studied the
conductance of a double degenerate (due to spin) quantum dot with Coulomb
repulsion and the e-ph interaction. Their numerical results also showed the
sideband peaks and the main peak related to the Coulomb repulsion, which was
decreased by the e-ph interaction. Kang \cite{kan} studied the boson
(vibron) assisted transport through a double-degenerate QD coupled to two 
{\it superconducting} leads and found multiple peaks in the I-V curves,
which originated from the singular BCS density of states and the phonon
sidebands. Ermakov \cite{erm} calculated the I-V curves of a four-fold
degenerate dot including both the onsite Coulomb and e-ph interactions. He
obtained a switching effect in the numerical I-V curves, similar to that in
the negative-$U$ Hubbard model discussed by us recently \cite{alebrawil}.
However, using a Hamiltonian averaged over the phonons, Ermakov missed all
phonon sidebands and obtained an unphysical population ($n=1)$ of each QD
state. More recently Gogolin and Komnik \cite{gog} have analyzed a
nondegenerate QD, $d=1,$ coupled with a single phonon mode. They found a
switching effect in the Born-Oppenheimer approximation similar to that in
our negative-$U$ Hubbard model, but surprisingly for a non-degenerate case
with $d=1$. However, we have to mention that the Born-Oppenheimer
approximation does not apply to a non-degenerate level, since there are no
``fast'' (compared to the characteristic phonon time $1/\omega _{0}$)
electron transitions within the dot. Despite differences in the models and
approximations, Refs. \cite{alebrawil,erm,gog} pointed to a novel mechanism
of the volatile molecular memory caused by the e-ph or any other attractive
electron correlations.

Here we develop the analytical theory of a {\it correlated} transport
through a degenerate molecule quantum dot (MQD) fully taking into account
both Coulomb and e-ph interactions. We show that the phonon sidebands
significantly modify the switching behavior of the I-V curves in comparison
with the negative-$U$ Hubbard model \cite{alebrawil}. Nevertheless, the
switching effect is robust. It shows up when the effective interaction of
polarons is attractive and the state of the dot is multiply degenerate, $d>2$%
.

\section{Steady current through MQD}

We apply the Landauer-type expression for the steady current through a
region of interacting electrons, derived by Meir and Wingreen \cite{meir} as 
(in units $\hbar=k_B=1$)
\begin{equation}
I(V)=-{e\over \pi}\int_{-\infty }^{\infty }d\omega \left[ f_{1}(\omega )-f_{2}(\omega )%
\right] {\rm Im Tr}\left[ \hat{\Gamma}(\omega )\hat{G}^{(1)}(\omega )\right] ,
\label{eq:Igen}
\end{equation}
where $f_{1(2)}(\omega )=\left\{ \exp [(\omega +\Delta 
\begin{array}{c}
- \\ 
(+)
\end{array}
eV/2)/T]+1\right\} ^{-1},$  $T$ is the temperature, $\Delta $ is the position of the lowest
unoccupied molecular level with respect to the chemical potential. $\hat{%
\Gamma}(\omega )$ depends on the density of states (DOS) in the leads and on
the hopping integrals connecting one-particle states in the left\ (1) and
the right (2) leads with the states in MQD, Fig.~1. This formula includes,
by means of the Fourier transform of the full molecular retarded Green's
function (GF), $\hat{G}^{(1)}(\omega ),$ the e-ph and Coulomb interactions
inside MQD and coupling to the leads. Since the leads are metallic,
electron-electron and e-ph interactions in the leads, and interactions of
electrons in the leads with electrons and phonons in MQD can be neglected.
We are interested in the tunneling near the conventional threshold, $%
eV=2\Delta $, Fig.1, within a voltage range about an effective attractive
potential $|U|$ caused by phonons/vibrons (see below). 
\begin{figure}[t]
\epsfxsize=2.8in \epsffile{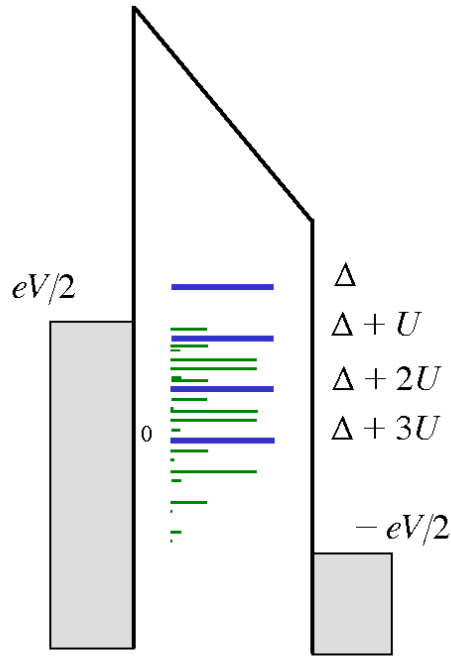}
\caption{ 
Schematic of the energy levels and phonon
side-bands for molecular quantum dot under bias voltage $V$
($eV/2\Delta=0.75$) with the coupling constant $\gamma^2=11/13$. The
level is assumed to be 4-fold degenerate $(d=4)$ with energies $\Delta+ rU$,
$r=0, \dots, (d-1)$ (thick bars). Thin bars show 
the vibron side-bands with the size of the bar proportional to the
weight of the particular contribution in the density of states (see
text) in the case of one vibron
with frequency $\omega_0/\Delta=0.2$ at $T=0$. Only the bands in the energy window
$(eV/2, -eV/2)$ (shown) will contribute to current at zero temperature.
 }
\label{fig:lad1}
\end{figure}

The attractive energy is the
difference of two large interactions, the Coulomb repulsion and the phonon
mediated attraction, of the order of $1{\rm eV}$ each. Hence, $|U|$ is of
the order of a few tens of one eV. We neglect the energy dependence of $%
\ \hat{\Gamma}(\omega )\approx \Gamma $ on this scale, and assume that the
coupling to the leads is weak, $\Gamma \ll $ $|U|.$ In this case $\hat{G}%
^{(1)}(\omega )$ does not depend on the leads. Moreover we assume that there
is a complete set of one-particle molecular states $\left| \mu \right\rangle 
$, where $\hat{G}^{(1)}(\omega )$ is diagonal. With these assumptions we can
reduce Eq.(\ref{eq:Igen}) to

\begin{equation}
I(V)=I_{0}\int_{-\infty }^{\infty }d\omega \left[ f_{1}(\omega
)-f_{2}(\omega )\right] \rho (\omega ),  \label{eq:Is}
\end{equation}
allowing for a transparent analysis of essential physics of the switching
phenomenon. Here $I_{0}=e\Gamma $ and the molecular DOS, $\rho (\omega ),$
is given by 
\begin{equation}
\rho (\omega )=-\frac{1}{\pi }\sum_{\mu }%
%TCIMACRO{\func{Im}}%
%BeginExpansion
\mathop{\rm Im}%
%EndExpansion
G_{\mu }(\omega ),
\end{equation}
where $G_{\mu }(\omega )$ is the Fourier transform of $G_{\mu }(t)=-i\theta
(t)\left\langle \left\{ c_{\mu }(t),c_{\mu }^{\dagger }\right\}
\right\rangle ,$ $\left\{ \cdots ,\cdots \right\} $ is the anticommutator, $%
c_{\mu }(t)=e^{iHt}c_{\mu }e^{-iHt},$ $\theta (t)=1$ for $t>0$ and zero
otherwise. We calculate $\rho (\omega )$ exactly in Section III in the
framework of the Hamiltonian, which includes both the Coulomb $U^C$ and e-ph
interactions as 
\begin{eqnarray}
H &=&\sum_{_{\mu }}\varepsilon _{_{\mu }}\hat{n}_{_{\mu }}+\frac{1}{2}%
\sum_{_{\mu }\neq \mu ^{\prime }}U^C_{\mu \mu ^{\prime }}\hat{n}_{_{\mu }}\hat{%
n}_{\mu ^{\prime }}  \nonumber \\
&&+\sum_{\mu ,q}\hat{n}_{_{\mu }}\omega _{q}(\gamma _{\mu
q}d_{q}+H.c.)+\sum_{q}\omega _{q}(d_{q}^{\dagger }d_{q}+1/2).
\end{eqnarray}
Here $\varepsilon _{_{\mu }}$ are one-particle molecular energy levels, $%
\hat{n}_{\mu }=c_{\mu }^{\dagger }c_{\mu }$ the occupation number operators, 
$c_{_{\mu }}$ and $d_{q}$ annihilates electrons and phonons, respectively, $%
\omega _{q}$ are the phonon (vibron) frequencies, and $\gamma _{\mu q}$ are
e-ph coupling constants ($q$ enumerates the vibron modes). This Hamiltonian
conserves the occupation numbers of molecular states $\hat{n}_{_{\mu }}$. Hence
it is compatible with Eq.(\ref{eq:Is}).

\section{MQD density of states}

We apply the canonical polaron unitary transformation $e^{S}$ \cite{fir},
integrating phonons out, as 
\begin{equation}
\tilde{H}=e^{S}He^{-S},
\end{equation}
where 
\begin{equation}
S=-\sum_{q,\mu }\hat{n}_{\mu }\left[ \gamma _{\mu q}d_{q}-H.c.\right]
\end{equation}
is such that $S^{\dagger }=-S.$ The electron and phonon operators are
transformed as 
\begin{equation}
\tilde{c}_{\mu }=c_{\mu }X_{\mu },
\end{equation}
and 
\begin{equation}
\tilde{d}_{q}=d_{q}-\sum_{\mu }\hat{n}_{\mu }\gamma _{\mu q}^{\ast },
\end{equation}
respectively. Here 
\[
X_{\mu }=\exp \left[ \sum_{q}\gamma _{\mu q}d_{q}-H.c.\right] . 
\]
The Lang-Firsov canonical transformation shifts ions to new equilibrium positions with
no effect on the phonon frequencies. The diagonalization is exact: 
\begin{equation}
\tilde{H}=\sum_{i}\tilde{\varepsilon}_{_{\mu }}\hat{n}_{\mu }+\sum_{q}\omega
_{q}(d_{q}^{\dagger }d_{q}+1/2)+{\frac{1}{{2}}}\sum_{\mu \neq \mu ^{\prime
}}U_{\mu \mu ^{\prime }}\hat{n}_{\mu }\hat{n}_{\mu ^{\prime }},
\end{equation}
where 
\begin{equation}
U_{\mu \mu ^{\prime }}\equiv U_{\mu \mu ^{\prime }}^{C}-2\sum_{q}\gamma
_{\mu q}^{\ast }\gamma _{\mu ^{\prime }q}\omega _{q}  \label{eq:Umm1}
\end{equation}
is the interaction of polarons comprising their interaction via molecular
deformations (vibrons) and non-vibron (e.g. Coulomb repulsion) $U_{\mu \mu
^{\prime }}^{C}$. To simplify the discussion, we shall assume, without
losing generality, that the Coulomb integrals do not depend on the orbital
index, i.e. $U_{\mu \mu ^{\prime }}=U.$

The molecular energy levels are shifted by the polaron level-shift due to a
deformation well created by polaron, 
\begin{equation}
\tilde{\varepsilon}_{_{\mu }}=\varepsilon _{_{\mu }}{-}\sum_{q}|\gamma _{\mu
q}|^{2}\omega _{q}.
\end{equation}
Applying the same transformation in the retarded GF we obtain 
\begin{eqnarray}
G_{\mu }(t) &=&-i\theta (t)\left\langle \left\{ c_{\mu }(t)X_{\mu
}(t),~c_{\mu }^{\dagger }X_{\mu }^{\dagger }\right\} \right\rangle \\
&=&-i\theta (t)[\left\langle c_{\mu }(t)c_{\mu }^{\dagger }\right\rangle
\left\langle X_{\mu }(t)X_{\mu }^{\dagger }\right\rangle  \nonumber \\
&&+\left\langle c_{\mu }^{\dagger }c_{\mu }(t)\right\rangle \left\langle
X_{\mu }^{\dagger }X_{\mu }(t)\right\rangle ],  \nonumber
\end{eqnarray}
where now electron and phonon operators are averaged over the quantum state
of the transformed Hamiltonian $\tilde{H}.$ There is no coupling between
polarons and vibrons in the transformed Hamiltonian, and the electron
and phonon averages are independent. The Heisenberg phonon operators evolve
as 
\begin{equation}
d_{q}(t)=d_{q}e^{-i\omega _{q}t},
\end{equation}
so that we find after thermodynamic averaging of the phonon correlator over
phonon occupation numbers (using the Weyl's identity for exponential
operators) 
\begin{eqnarray}
&&\left\langle X_{\mu }(t)X_{\mu }^{\dagger }\right\rangle  \nonumber \\
&=&\exp \left[ \sum_{q}\frac{|\gamma _{\mu q}|^{2}}{\sinh \frac{\beta \omega
_{q}}{2}}\left[ \cos \left( \omega t+i\frac{\beta \omega _{q}}{2}\right)
-\cosh \frac{\beta \omega _{q}}{2}\right] \right],  \label{eq:XXT}
\end{eqnarray}
where $\beta=1/T$.
Repeating the calculations for $\left\langle X_{\mu }^{\dagger }X_{\mu
}(t)\right\rangle $ we find a simple useful relation 
\begin{equation}
\left\langle X_{\mu }^{\dagger }X_{\mu }(t)\right\rangle =\left\langle
X_{\mu }(t)X_{\mu }^{\dagger }\right\rangle ^{\ast }.  \label{eq:X+X}
\end{equation}
At low temperatures $T\ll \omega _{q}$ the phonon correlator (\ref{eq:XXT})
simplifies to 
\begin{equation}
\left\langle X_{\mu }(t)X_{\mu }^{\dagger }\right\rangle =\exp \left[ \sum_{%
{\bf q}}|\gamma _{\mu q}|^{2}(e^{^{-i\omega _{q}t}}-1)\right] .
\label{eq:X1}
\end{equation}
Next, we introduce the multi-particle GFs, which will necessarily appear in
the equations of motion for $\left\langle c_{\mu }(t)c_{\mu }^{\dagger
}\right\rangle $, as 
\begin{equation}
G_{\mu }^{(r,+)}(t)\equiv -i\theta (t)\sum_{\mu _{1}\neq \mu _{2}\neq ...\mu
}\left\langle c_{\mu }(t)c_{\mu }^{\dagger }\prod_{i=1}^{r-1}\hat{n}_{\mu
_{i}}\right\rangle ,
\end{equation}
and 
\begin{equation}
G_{\mu }^{(r,-)}(t)\equiv -i\theta (t)\sum_{\mu _{1}\neq \mu _{2}\neq ...\mu
}\left\langle c_{\mu }^{\dagger }c_{\mu }(t)\prod_{i=1}^{r-1}\hat{n}_{\mu
_{i}}\right\rangle . 
\end{equation}
Then, using the equation of motion for the Heisenberg polaron operator, 
\begin{equation}
i\frac{dc_{\mu }}{dt}=\left( \tilde{\varepsilon}_{_{\mu }}+U\sum_{\mu
^{\prime }(\neq \mu )}\hat{n}_{\mu ^{\prime }}\right) c_{_{\mu }},
\end{equation}
we derive the following equations for the multi-particle GFs, 
\begin{eqnarray}
i\frac{dG_{\mu }^{(r,+)}(t)}{dt}=\delta (t)(1-n_{\mu })\sum_{\mu _{1}\neq
\mu _{2}\neq ...\mu }\prod_{i=1}^{r-1}n_{\mu _{i}}  \nonumber \\
+[\tilde{\varepsilon}_{_{\mu }}+(r-1)U]G_{\mu }^{(r,+)}(t)+UG_{\mu
}^{(r+1,+)}(t),
\end{eqnarray}
and 
\begin{eqnarray}
i\frac{dG_{\mu }^{(r,-)}(t)}{dt}=\delta (t)n_{\mu }\sum_{\mu _{1}\neq \mu
_{2}\neq ...\mu }\prod_{i=1}^{r-1}n_{\mu _{i}}  \nonumber \\
+[\tilde{\varepsilon}_{_{\mu }}+(r-1)U]G_{\mu }^{(r,-)}(t)+UG_{\mu
}^{(r+1,-)}(t),
\end{eqnarray}
where $n_{\mu }=\left\langle c_{\mu }^{\dagger }c_{\mu }\right\rangle $ is
the expectation number of electrons on the molecular level $\mu $.

We can readily solve this set of coupled equations for MQD with one $d$-fold
degenerate energy level
and with the e-ph coupling $\gamma _{\mu q}=\gamma _{q},$
which does not break the degeneracy. At zero bias voltage the empty level
will lie by some energy $\Delta $ above the Fermi levels of the
electrodes (Fig.~1).
Assuming that $n_{\mu }=n,$ Fourier transformation of the set yields 
\begin{equation}
G_{\mu }^{(1,+)}(\omega )=(1-n)\sum_{r=0}^{d-1}\frac{Z_{r}(n)}{\omega
-rU+i\delta },  \label{eq:G1+}
\end{equation}
\begin{equation}
G_{\mu }^{(1,-)}(\omega )=n\sum_{r=0}^{d-1}\frac{Z_{r}(n)}{\omega
-rU+i\delta }  \label{eq:G1-}
\end{equation}
where $\delta =+0$, and 
\begin{equation}
Z_{r}(n)=\frac{(d-1)!}{r!(d-1-r)!}n^{r}(1-n)^{d-1-r}. 
\end{equation}

\subsection{MQD\ Green's function at low temperatures ($T\ll \protect\omega
_{q})$}

It is the easiest to find the total Green's function at low temperatures.
Indeed, transforming back to real time and using Eqs.(\ref{eq:G1+}), (\ref
{eq:G1-}) and (\ref{eq:X1}), (\ref{eq:X+X}) we arrive at

\begin{eqnarray}
G_{\mu }(t) &=&-i\theta (t){\cal Z}_{0}\sum_{r=0}^{d-1}Z_{r}(n)e^{-irUt} 
\nonumber \\
&&\times \biggl[(1-n)\exp \left( \sum_{q}|\gamma _{\mu q}|^{2}e^{-i\omega
_{q}t}\right)  \nonumber \\
&&+n\exp \left( \sum_{q}|\gamma _{\mu q}|^{2}e^{i\omega _{q}t}\right) \biggr]%
,  \label{eq:Gt}
\end{eqnarray}
where ${\cal Z}_{0}=\exp \left[ -\sum_{{\bf q}}|\gamma _{q}|^{2}\right] $.
This is an exact solution with respect to correlations and e-ph interactions
which satisfies all sum rules. Expanding the exponents in the brackets we
find its Fourier component as 
\begin{eqnarray}
&&G_{\mu }(\omega ) ={\cal Z}_{0}\sum_{r=0}^{d-1}Z_{r}(n)\biggl[\frac{1}{%
\omega -rU+i\delta }  \nonumber \\
&&+\sum_{l=1}^{\infty }\frac{1}{l!}\sum_{q_{1},...q_{l}}|\gamma
_{q_{1}}...\gamma _{q_{l}}|^{2}\biggl(\frac{1-n}{\omega
-rU-\sum_{k=1}^{l}\omega _{v_{k}}+i\delta }  \nonumber \\
&&+\frac{n}{\omega -rU+\sum_{k=1}^{l}\omega _{v_{k}}+i\delta }\biggr)\biggr]
\label{eq:Gw}
\end{eqnarray}
If the e-ph interaction is weak, $|\gamma _{q}|\ll 1$, the essential
contribution comes only from the first (phonon-less) term, and we recover the
result of the Hubbard model \cite{alebrawil}

\begin{equation}
G_{\mu }(\omega )=\sum_{r=0}^{d-1}\frac{Z_{r}(n)}{\omega -rU+i\delta }.
\end{equation}
At finite $|\gamma _{q}|\gtrsim 1$, the phonon side-bands become important
in Eq.(\ref{eq:Gw}), which is obviously in the form of the multi-polaron
expansion. If one neglects the correlations, $U=0,$ a standard polaron GF 
\cite{alemot} is recovered:

\begin{eqnarray}
G_{\mu }(\omega )={\cal Z}_{0}\biggl[\frac{1}{\omega +i\delta }%
+\sum_{l=1}^{\infty }\frac{1}{l!}\sum_{q_{1},...q_{l}}|\gamma
_{q_{1}}...\gamma _{q_{l}}|^{2}  \nonumber \\
\times \left( \frac{1-n}{\omega -\sum_{k=1}^{l}\omega _{v_{k}}+i\delta }+%
\frac{n}{\omega +\sum_{k=1}^{l}\omega _{v_{k}}+i\delta }\right) \biggr],
\label{eq:GnoU}
\end{eqnarray}
where we have applied the sum rule 
\begin{equation}
\sum_{r=0}^{d-1}Z_{r}(n)=1.
\end{equation}

\subsection{MQD Green's function at finite temperatures, single vibron mode}

By applying the same method, as in the case of $T=0,$ and going over
back to real 
time with the  use of Eqs.(\ref{eq:G1+}), (\ref{eq:G1-}) and
(\ref{eq:X1}), (\ref {eq:X+X}) we arrive at

\begin{eqnarray}
G_{\mu }(t) &=&-i\theta (t){\cal Z}\sum_{r=0}^{d-1}Z_{r}(n)e^{-irUt} 
\nonumber \\
&&\times \biggl[(1-n)\exp \left( \sum_{q}\frac{|\gamma _{\mu q}|^{2}}{\sinh 
\frac{\beta \omega _{q}}{2}}\cos \left( \omega t+i\frac{\beta \omega _{q}}{2}%
\right) \right)  \nonumber \\
&&+n\exp \left( \sum_{q}\frac{|\gamma _{\mu q}|^{2}}{\sinh \frac{\beta
\omega _{q}}{2}}\cos \left( \omega t-i\frac{\beta \omega _{q}}{2}\right)
\right) \biggr],  \label{eq:GtT}
\end{eqnarray}
where now 
\begin{equation}
{\cal Z}=\exp \left[ -\sum_{{\bf q}}|\gamma _{q}|^{2}\coth \frac{\beta
\omega _{q}}{2}\right] .
\end{equation}

In approximation, where we retain a coupling to a single mode with the
characteristic frequency $\omega _{0}$ and $\gamma _{q}\equiv \gamma $, we
can expand the exponents in the temporal Green's function (\ref{eq:GtT}) in
powers of $\exp \left( \omega t+i\frac{\beta \omega _{0}}{2}\right) $. It is
then trivial to find the Green's function in the frequency domain as 
\begin{eqnarray}
&&G_{\mu }(\omega ) ={\cal Z}\sum_{r=0}^{d-1}Z_{r}(n)\sum_{l=0}^{\infty
}I_{l}\left( \xi \right)  \nonumber \\
&&\times \biggl[e^{\frac{\beta \omega _{0}l}{2}}\left( \frac{1-n}{\omega
-rU-l\omega _{0}+i\delta }+\frac{n}{\omega -rU+l\omega _{0}+i\delta }\right)
\nonumber \\
&&+(1-\delta _{l0})e^{-\frac{\beta \omega _{0}l}{2}}\bigl(\frac{1-n}{\omega
-rU+l\omega _{0}+i\delta }  \nonumber \\
&&+\frac{n}{\omega -rU-l\omega _{0}+i\delta }\bigr))\biggr],  \label{eq:GwT}
\end{eqnarray}
where $\xi =|\gamma |^{2}/\sinh \frac{\beta \omega _{0}}{2},$ $I_{l}\left(
\xi \right) $ is the modified Bessel function, and $\delta _{lk}$ is the
Kroneker symbol. At low temperatures, where $\beta \omega _{0}\gg 1,$ $\xi
\ll 1$ and $I_{l}\left( \xi \right) \approx \left( \frac{\xi }{2}\right)
^{l}/l!,$ this expression gives (\ref{eq:Gw}) in the form 
\begin{eqnarray}
G_{\mu }(\omega ) &=&{\cal Z}_{0}\sum_{r=0}^{d-1}Z_{r}(n)\sum_{l=0}^{\infty }%
\frac{|\gamma |^{2l}}{l!}\biggl(\frac{1-n}{\omega -rU-l\omega _{0}+i\delta }
\nonumber \\
&&+\frac{n}{\omega -rU+l\omega _{0}+i\delta }\biggr).  \label{eq:GwT0sm}
\end{eqnarray}

The {\em molecular DOS} is readily found as an imaginary part of Eq.(\ref
{eq:GwT}): 
\begin{eqnarray}
&&\rho (\omega ) ={\cal Z}d\sum_{r=0}^{d-1}Z_{r}(n)\sum_{l=0}^{\infty
}I_{l}\left( \xi \right)  \nonumber \\
&&\times \biggl[e^{\frac{\beta \omega _{0}l}{2}}\left[ (1-n)\delta (\omega
-rU-l\omega _{0})+n\delta (\omega -rU+l\omega _{0})\right]  \nonumber \\
&&+(1-\delta _{l0})e^{-\frac{\beta \omega _{0}l}{2}}[n\delta (\omega
-rU-l\omega _{0})   \nonumber \\
&&+(1-n)\delta (\omega -rU+l\omega _{0})]\biggr].\label{eq:rho}
\end{eqnarray}
The important feature of the DOS, Eq.(\ref{eq:rho}), is its nonlinear
dependence on the occupation number $n,$ which leads to the switching effect
and hysteresis in the I-V characteristics for $d>2$, as will be shown below.
It  contains full information about all possible correlation and
inelastic effects in transport, in particular, all the vibron-assisted
tunneling processes (Fig. 2). 
\begin{figure}[t]
\epsfxsize=2.8in \epsffile{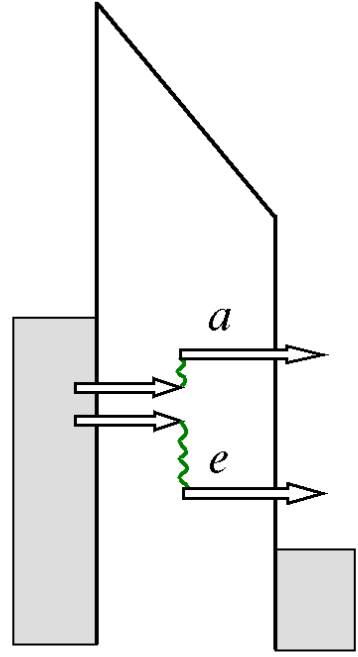}
\caption{ 
Schematic of the inelastic processes 
assisting tunneling through the molecular quantum dot (see Fig. 1).
The tunneling through coupled electron-vibron system may proceed with
the emission (process $e$) or absorption (process $a$) of the
vibrons. The absorption is possible only at non-zero temperatures.
 }
\label{fig:lad2}
\end{figure}

\section{Nonlinear rate equation and switching}

Generally, the electron density $n_{\mu }$ obeys an infinite set of rate
equations for many-particle GFs which can be derived in the framework of a
tunneling Hamiltonian including correlations \cite{alebrawil}. In the case
of MQD only weakly coupled with leads one can apply the Fermi-Dirac golden
rule to obtain an equation for $n.$ Equating incoming and outgoing numbers
of electrons in MQD per unit time we obtain the self-consistent equation for
the level occupation $n$ as 
\begin{eqnarray}
&&(1-n)\int_{-\infty }^{\infty }d\omega \left\{ \Gamma _{1}f_{1}(\omega
)+\Gamma _{2}f_{2}(\omega )\right\} \rho (\omega )  \nonumber \\
&&-n\int_{-\infty }^{\infty }d\omega \left\{ \Gamma _{1}[1-f_{1}(\omega
)]+\Gamma _{2}[1-f_{2}(\omega )]\right\} \rho (\omega ) =0  \label{eq:neq}
\end{eqnarray}
where $\Gamma _{1(2)}$ are the transition rates from left (right) leads to MQD.
Taking into account that $\int_{-\infty }^{\infty }\rho (\omega )=d$, Eq.(%
\ref{eq:neq}) for the symmetric leads, $\Gamma _{1}=\Gamma _{2},$ 
reduces to
\begin{equation}
2nd=\int d\omega \rho \left( \omega \right) \left( f_{1}+f_{2}\right)
, 
\end{equation}
which automatically satisfies $0\leq n \leq 1$.
Explicitly, the self-consistent equation for the occupation number is
\begin{equation}
n=\frac{1}{2}\sum_{r=0}^{d-1}Z_{r}(n)[na_{r}+(1-n)b_{r}],  \label{eq:neq1}
\end{equation}
where 
\begin{eqnarray}
a_{r} &=&{\cal Z}\sum_{l=0}^{\infty }I_{l}\left( \xi \right) \biggr(e^{\frac{%
\beta \omega _{0}l}{2}}[f_{1}(rU-l\omega _{0})+f_{2}(rU-l\omega _{0})] 
\nonumber \\
&&+(1-\delta _{l0})e^{-\frac{\beta \omega _{0}l}{2}}[f_{1}(rU+l\omega
_{0})+f_{2}(rU+l\omega _{0})]\biggr),  \label{eq:a} \\
b_{r} &=&{\cal Z}\sum_{l=0}^{\infty }I_{l}\left( \xi \right) \biggr(e^{\frac{%
\beta \omega _{0}l}{2}}[f_{1}(rU+l\omega _{0})+f_{2}(rU+l\omega _{0})] 
\nonumber \\
&&+(1-\delta _{l0})e^{-\frac{\beta \omega _{0}l}{2}}[f_{1}(rU-l\omega
_{0})+f_{2}(rU-l\omega _{0})]\biggr).  \label{eq:b}
\end{eqnarray}
The current is expressed as 
\begin{equation}
j\equiv \frac{I(V)}{dI_{0}}=\sum_{r=0}^{d-1}Z_{r}(n)[na_{r}^{\prime
}+(1-n)b_{r}^{\prime }],
\end{equation}
where 
\begin{eqnarray}
a_{r}^{\prime } &=&{\cal Z}\sum_{l=0}^{\infty }I_{l}\left( \xi \right) %
\biggr(e^{\frac{\beta \omega _{0}l}{2}}[f_{1}(rU-l\omega
_{0})-f_{2}(rU-l\omega _{0})]  \nonumber \\
&&+(1-\delta _{l0})e^{-\frac{\beta \omega _{0}l}{2}}[f_{1}(rU+l\omega
_{0})-f_{2}(rU+l\omega _{0})]\biggr),  \label{eq:at} \\
b_{r}^{\prime } &=&{\cal Z}\sum_{l=0}^{\infty }I_{l}\left( \xi \right) %
\biggr(e^{\frac{\beta \omega _{0}l}{2}}[f_{1}(rU+l\omega
_{0})+f_{2}(rU+l\omega _{0})]  \nonumber \\
&&+(1-\delta _{l0})e^{-\frac{\beta \omega _{0}l}{2}}[f_{1}(rU-l\omega
_{0})+f_{2}(rU-l\omega _{0})]\biggr).  \label{eq:bt}
\end{eqnarray}

Let us analyze the I-V curves for $d=1$, $2$, and $4.$

\subsection{Absence of switching for nondegenerate and two--fold degenerate
MQD}

There is one term in the sum over $r$, $r=0$ with $Z_{0}(n)=1,$ if $d=1.$
Hence there is only one solution of the rate equation (\ref{eq:neq1}) 
\begin{equation}
n=\frac{b_{0}}{2+b_{0}-a_{0}},
\end{equation}
and the current is single-valued at any voltage 
\begin{equation}
j=\frac{2b_{0}^{\prime }+a_{0}^{\prime }b_{0}-a_{0}b_{0}^{\prime }}{%
2+b_{0}-a_{0}}.
\end{equation}
This is an exact result, which is valid for any e-ph coupling and any phonon
frequency. We have to conclude that there is no switching of a nondegenerate
MQD. The opposite conclusion reached in Ref.\cite{gog} might be due to the
Born-Oppenheimer (static) approximation used by Gogolin and Komnik. In fact, 
the Born-Oppenheimer approximation does not apply to the non-degenerate
level model, since there are no ``fast'', compared to phonon times $1/\omega
_{0},$ electron transitions within the ``molecule''.

In the case of a double-degenerate MQD, $d=2,$ there are two terms, which
contribute to the sum over $r$, with $Z_{0}(n)=1-n$ and $Z_{1}(n)=n.$ The
rate equation becomes a quadratic one 
\begin{equation}
n^{2}(a_{0}+a_{1}-b_{0}-b_{1})+n(2-a_{0}+b_{0}-b_{1})-b_{0}=0.
\end{equation}
with two solutions, 
\begin{eqnarray}
&&n_{1.2} =-\frac{2-a_{0}+b_{0}-b_{1}}{2(a_{0}+a_{1}-b_{0}-b_{1})}  \nonumber
\\
&&\pm \left[ \frac{(2-a_{0}+b_{0}-b_{1})^{2}}{4((a_{0}+a_{1}-b_{0}-b_{1})^{2}%
}+\frac{b_{0}}{a_{0}+a_{1}-b_{0}-b_{1}}\right] ^{1/2}.  \label{eq:n12}
\end{eqnarray}
However, one of them is negative because $0<b_{r}<a_{r}<1$ for any
temperature and voltage. Therefore, we conclude that there is only one
physical population of MQD, and the current is also single-valued at any
voltage and temperature, in agreement with the Hubbard model \cite{alebrawil}%
.

\subsection{Switching of four-fold degenerate MQD}

In this case the rate equation is of the fourth power in $n,$%
\begin{eqnarray}
2n &=&(1-n)^{3}[na_{0}+(1-n)b_{0}]  \nonumber \\
&&+3n(1-n)^{2}[na_{1}+(1-n)b_{1}]  \nonumber \\
&&+3n^{2}(1-n)[na_{2}+(1-n)b_{2}]  \nonumber \\
&&+n^{3}[na_{3}+(1-n)b_{3}].  \label{eq:nab}
\end{eqnarray}
In the limit $|\gamma |\ll 1$ we have $b_{r}=a_{r}$, ${\cal Z}=1,$ the
remaining interaction is $U=U^{C}$ [see Eq.(\ref{eq:Umm1})] and Eq.(\ref
{eq:nab}) is reduced to 
\begin{eqnarray}
2n &=&(1-n)^{3}b_{0}+3n(1-n)^{2}b_{1} \\
&&+3n^{2}(1-n)b_{2}+n^{3}b_{3}.  \nonumber
\end{eqnarray}
If we assume now that the non-vibron interaction $U^{C}$ is negative, for
example due to valence fluctuations, then we recover the negative-$U$ model 
\cite{alebrawil}, and the kinetic equation is reduced to 
\begin{equation}
2n=1-(1-n)^{3}  \label{eq:ncub}
\end{equation}
in the voltage range $\Delta -|U|<eV/2<\Delta $ and $T=0$ because $b_{0}=0$
and $b_{1}=b_{2}=b_{3}=1$ there, if $|U|<\Delta /2$. The
current is simplified as 
\begin{equation}
j=2n.
\end{equation}
Equation (\ref{eq:ncub}) has two physical roots, $n=0$ and $n=(3-5^{1/2})/2\approx
0.38$. Hence we obtain two stationary states of MQD with low (zero at $T=0$)
and high current, $j\approx 0.76$ for the same voltage as we discussed
earlier in Ref. \cite{alebrawil}. The current-voltage characteristics show a
hysteretic behavior for $d=4.$ When the voltage increases from zero, 4-fold
degenerate MQD remains in a low-current state until the threshold $%
eV_{2}/2=\Delta $ is reached. Remarkably, when the voltage {\em decreases}
from the value above the threshold $V_{2}$, the molecule remains in the
high-current state down to the voltage $eV_{1}/2=\Delta -|U|$ well below the
threshold $V_{2}$. This is a correlation mechanism of electronic molecular
switching without retardation. Therefore, the negative-$U$ degenerate molecular dot
possesses the volatile memory originating from the many-particle attractive
correlations.
\begin{figure}[t]
\epsfxsize=3in \epsffile{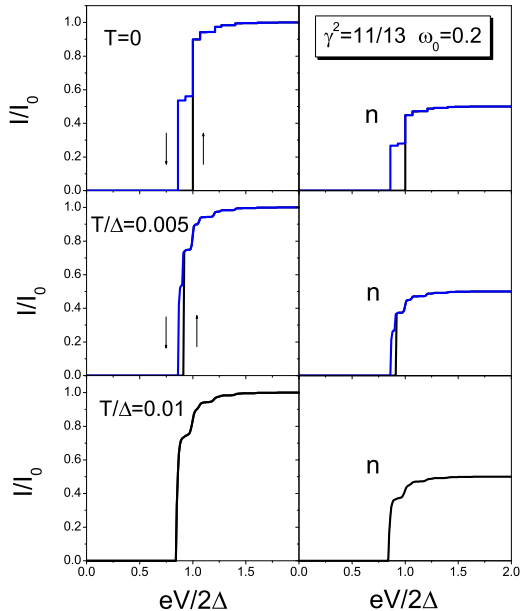}
\caption{ The I-V curves for tunneling through molecular quantum dot
(Fig. 1) with the electron-vibron coupling constant $\gamma^2=11/13$
and $\omega_0/\Delta = 0.2$. The up arrows show that the current picks up
at some voltage when it is biased, and then drops at lower
voltage when the bias is being reduced. The bias dependence of current basically
repeats the shape of the level occupation $n$ (right column). Steps on
the curve correspond to the changing population of the phonon side-bands, which
are shown in Fig. 1. The current hysteresis persist up to some
critical temperature, which is low, $T/\Delta \approx 0.01$.
 }
\label{fig:g1113}
\end{figure}
\begin{figure}[t]
\epsfxsize=3in \epsffile{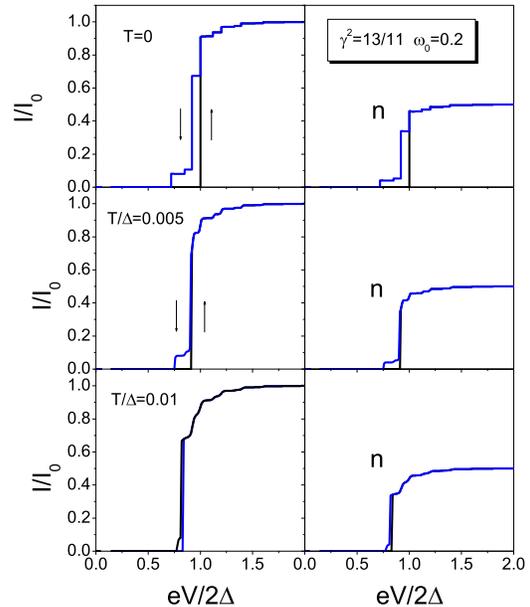}
\caption{ The I-V curves for tunneling through molecular quantum dot
with the electron-vibron coupling constant $\gamma^2=13/11$, which is
slightly larger than in Fig.~3. The I-V curves
change substantially: the current pickup shifts to lower bias voltages
and the curves show substantial change of shape compared to
Fig. 3. The hysteresis persists to 
slightly higher temperatures, although also small, as compared to the
previous case.
 }
\label{fig:g1311}
\end{figure}

The e-ph coupling results in the phonon sidebands of DOS, which are fully
taken into account in Eq.(\ref{eq:neq}) for the self-consistent occupation
of the molecular level $n.$ In the case of the Coulomb repulsion and the
electron-vibron coupling the effective interaction will be attractive, if 
\begin{equation}
|\gamma |\geq \left( \frac{U^{C}}{2\omega _{0}}\right) ^{1/2}.
\end{equation}

There is an important difference between switching with vibron-mediated
electron-electron attraction and the negative-$U$ model. We show the
numerical results for $\omega _{0}=0.2$ (in units of $\Delta ,$ as all the
energies in the problem) and $U^{C}=0$ for two values of the coupling
constant, $\gamma ^{2}=11/13$ (Fig. 3)\ and $\gamma ^{2}=13/11$ (Fig. 4).
This case formally corresponds to $U=-2\gamma ^{2}\omega _{0}\approx -0.4$,
i.e. close to the same value of the attraction as we have used in the
negative-$U$ model \cite{alebrawil} (we selected those values of
$\gamma^2$ to avoid
accidental commensurability of a ladder of the correlated level energies
separated by $U$ and generated by them phonon side-band ladders 
with the step $\omega_{0}) $. In the negative-$U$\ case the threshold voltages were $%
eV_{1}/2\Delta =1-0.4=0.6$ and $eV_{2}/2\Delta =1.$ However, in the vibron
case the threshold for the onset of bistability corresponds to larger
voltage bias compared to negative-$U$ case (at $eV/2\Delta =0.86$ for $%
\gamma ^{2}=11/13$ and $\omega _{0}=0.2$). The I-V in the vibron case is
much more complex too:\ as one can see from Figs. 3 and 4, the current
discontinuity at the threshold strongly depends on the value of the e-ph
coupling constant.  The
inelastic tunneling processes through the level, accompanied by
emission/absorption of the phonons (Fig. 2), manifest themselves as steps on the I-V
curve, Figs. 3 and 4. Those steps are generated by the phonon side-bands
originating from correlated levels on the dot with the energies $\Delta ,$ $%
\Delta +U,$ ..., $\Delta +(d-1)U.$ Since $\omega _{0}$ is not generally
commensurate with $U,$ we obtain pretty irregular picture of the steps on
I-V\ curve. This comes as no surprise, since the kinetic equation (\ref
{eq:nab})\ is much more complex compared to the one in the negative-U\ case,
cf. (\ref{eq:ncub}). The level occupation approaches the limiting value $%
n=0.5$ at large bias voltages, Figs. 3 and 4.

The bistability region shrinks down with temperature. In the specific
example of the negative-$U$ model with $U/\Delta =-0.4$ the bistability is
over at $T/\Delta \approx 0.1$ \cite{alebrawil}$.$ Importantly, in the
vibron case this happens at much lower temperatures. Indeed, the hysteresis
loop almost closes at $T/\Delta =0.01$. The critical temperature, below
which the current bistability exists in the vibron case, is suppressed by
about an order of magnitude compared to the negative-$U$ case.
At finite temperatures, the overall I-V curve shows the smoothed out
steps at the bias voltages coinciding with the voltage where the vibrons are
emitted/absorbed as at $T=0$ (Figs.3, 4).

In conclusion, we have developed the multi-polaron theory of tunneling
through a molecular quantum dot (MQD) taking phonon sidebands and strong
electron correlations into account. The degenerate MQD with strong
electron-vibron coupling shows a hysteretic volatile memory if the degeneracy
of the molecular level is larger than two, $d>2.$ The hysteretic behavior
strongly depends on electron-vibron coupling and characteristic vibron
frequencies. The current bistability vanishes above some critical
temperature. It would be very interesting to look for an experimental
realization of the model, possibly in a system containing a certain
conjugated central part, which exhibits the attractive
correlations of carriers with large degeneracy $d>2.$

This work has been partly supported by DARPA. The authors acknowledge useful
discussions with V.V. Osipov and R.S. Williams.


\begin{references}
\bibitem{kas}  M.A. Kastner, Rev. Mod. Phys. {\bf 64}, 849 (1992) and
references therein.

\bibitem{but}  M. B\"{u}ttiker, Phys. Rev. Lett. {\bf 57}, 1761 (1986).

\bibitem{and}  P.W. Anderson, E. Abrahams, D.S. Fisher, and D.J. Thouless,
Phys. Rev. B{\bf 22}, 3519 (1980).

\bibitem{lee}  P.A. Lee and D.S. Fisher, Phys. Rev. Lett. {\bf 47}, 882
(1981).

\bibitem{sha}  B. Shapiro, Phys. Rev. Lett. {\bf 48}, 823 (1982).

\bibitem{lik}  D.V. Averin, A.N. Korotkov, and K.K. Likharev, Phys. Rev. B%
{\bf 44}, 6199 (1991).

\bibitem{meir}  Y. Meir and N.S. Wingreen, Phys. Rev. Lett. {\bf 68}, 2512
(1992).

\bibitem{her}  S. Hershfield, J.H. Davies, and J.W. Wilkins, Phys. Rev. B%
{\bf 46}, 7046 (1992).

\bibitem{par}  J. Park, A.N. Pasupathy, J.I. Goldsmith, C.Chang, Y. Yaish,
J.R. Retta, M. Rinkoski, J.P. Sethna, H.D. Abru\~{n}a, P.L. McEuen, and D.C.
Ralph, Nature (London) {\bf 417}, 722 (2000).

\bibitem{het}  M.H. Hettler, H. Schoeller, and W. Wenzel, cond-mat/0011047.

\bibitem{lehn90}  J.-M. Lehn, Angew. Chem. Int. Ed. Engl. {\bf 29}, 1304
(1990).

\bibitem{tour00}  J.M. Tour, Acc. Chem. Res. {\bf 33}, 791 (2000); J.M. Tour 
{\it et al}., J. Am. Chem. Soc. {\bf 117}, 9529 (1995).

\bibitem{mark98}  A. Aviram and M. Ratner, Eds., {\it Molecular
Electronics:\ Science and Technology} (Ann. N.Y. Acad. Sci., New York, 1998).

\bibitem{pat99}  C.P. Collier {\it et al.}, Science {\bf 285}, 391 (1999);
J. Chen {\it et al.}, Science {\bf 286}, 1550 (1999); D.I. Gittins {\it et
al.}, Nature (London) {\bf 408}, 677 (2000); H. X. He , T.J. Tao, L.A.
Nagahara, I. Amlani and R. Tsui (unpublished).

\bibitem{exp}  D. Stewart {\it et al.} (unpublished).

\bibitem{alebrawil}  A.S. Alexandrov, A.M. Bratkovsky, and R.S. Williams,
Phys. Rev. B (to be published); cond-mat/0204387.

\bibitem{zhit02} N.B. Zhitenev, H. Meng, and Z. Bao,
Phys. Rev. Lett. {\bf 88}, 226801 (2002).

\bibitem{win2}  L.I. Glazman and R.I. Shekhter, Zh. Eksp. Teor. Fiz. {\bf 94}%
, 292 (1987) [Sov. Phys. JETP {\bf 67}, 163 (1988)]; N.S. Wingreen, K.W.
Jacobsen, and J.W. Wilkins, Phys. Rev. B{\bf 17}, 11834 (1989).

\bibitem{li}  Xi Li, H. Chen, and S. Zhou, Phys. Rev. B{\bf 52}, 12202
(1995).

\bibitem{kan}  K. Kang, Phys. Rev. B{\bf 57}, 11891 (1998).

\bibitem{erm}  V.N. Ermakov, Physica E{\bf 8}, 99 (2000).

\bibitem{ven}  M. Di Ventra, S.-G. Kim, S.T. Pantelides, and N.D. Lang,
Phys. Rev. Lett. {\bf 86}, 288 (2001).

\bibitem{fis}  N. Ness, S. A. Shevlin, and A. J. Fisher, Phys. Rev. B {\bf 63%
}, 125422 (2001).

\bibitem{lun}  U. Lundin and R.H. McKenzie, cond-mat/0203548.

\bibitem{gog}  A.O. Gogolin and A. Komnik, cond-mat/0207513.

\bibitem{fir}  I.G. Lang and Yu. A. Firsov, Zh. Eksp. Teor. Fiz. {\bf 43},
1843 (1962) [Sov. Phys. JETP {\bf 16}, 1301 (1963)].

\bibitem{alemot}  A. S. Alexandrov and N. F. Mott, {\it Polarons and
Bipolarons} (World Scientific, Singapore, 1995).
\end{references}
\end{document}